\documentclass{aa}
\usepackage{graphicx}
\begin{document}
\title{Nuclear composition and heating\\  in accreting neutron-star crusts}

\author{P. Haensel
 \and
 J.L. Zdunik
}
 \institute{N. Copernicus Astronomical
Center, Polish
           Academy of Sciences, Bartycka 18, PL-00-716 Warszawa, Poland
           {\em haensel@camk.edu.pl}}
 \institute{N. Copernicus Astronomical
Center, Polish
           Academy of Sciences, Bartycka 18, PL-00-716 Warszawa, Poland\\
%
{\tt   haensel@camk.edu.pl, jlz@camk.edu.pl}
}
\offprints{P. Haensel}
\date{Received  / Accepted }
\abstract{ Nuclear reactions in accreting
neutron-star crusts and the heat release accompanying them
are studied, under different assumptions concerning the
composition of the outermost layer formed of the ashes of X-ray bursts. Particular
examples  of ashes containing nuclides with $A\simeq 90-110$ are considered and
compared with a standard $A=56$ case. In all cases, evolution of a crust shell is
followed from $10^8~{\rm g~cm^{-3}}$ to a few times $10^{13}~{\rm g~cm^{-3}}$. The
total crustal heating produced  in the non-equilibrium processes in the accreting crust
is  $1.1-1.5~{\rm MeV}$ per one accreted nucleon. The composition of the accreted crust at
 densities exceeding the threshold for the pycnonuclear fusion
 ($\rho>10^{12}~{\rm g~cm^{-3}}$) is  essentially independent of the assumed
 initial composition of the  X-ray burst ashes.
\keywords{dense matter -- equation
 of state -- stars: neutron
-- stars :general -- X-rays: bursts -- X-ray: binaries -- nuclear
reactions} }
\titlerunning{Composition and heating in accreting neutron-star crusts}
\authorrunning{P. Haensel and J.L. Zdunik}
\maketitle
\section{Introduction}
\label{sect:introduction}
Accretion of matter onto neutron stars in close binary systems plays crucial role in
many objects studied by the X-ray astronomy, such as X-ray bursters, X-ray pulsars,
and transient X-ray sources. Particular attention was focused recently on soft X-ray
transients (SXRTs) in quiescence, when the accretion from a disk formed of matter
flowing from the low-mass companion star is switched off or strongly suppressed. The
nature of the X-ray radiation during quiescence is still a matter of debate.
Typically, the quiescent emission is much higher than the expected one for an old
cooling neutron star. It has been suggested that this is due to the fact that the
interiors of neutron stars in SXRTs are heated-up,  during relatively short periods of
accretion and bursting, by the non-equilibrium processes associated with nuclear
reactions taking place in the deep layers of the crust (Brown et al. 1998). This
additional {\it crustal heating}, combined with appropriate models of neutron-star
atmosphere and interior, can be used to explain observations of SXRTs in quiescence.
Dependence of the luminosity in quiescence on the structure of neutron-star core, and
particularly on the rate of neutrino cooling,  opened a new possibility of exploring
the internal structure and equation of state of neutron stars via confrontation of
theoretical models with observations of quiescent SQRTs (see, e.g., Ushomirsky \&
Rutledge 2001, Colpi et al. 2001, Rutledge et al. 2002, Brown et al. 2002, and
references therein).

The crustal heating used in the SQRTs models was described using the results of the
calculations of  Haensel \& Zdunik (1990)(hereafter referred to as HZ). In the HZ
model, an outer layer of neutron star, formed  in X-ray bursts, is composed of
$^{56}{\rm Fe}$. Such a composition, corresponding to the maximum nuclear binding, was
predicted by the standard models of the thermonuclear X-ray bursts of that time. Then,
the $^{56}{\rm Fe}$ layer is sinking in the neutron star interior under the weight of
accreted matter. Under increasing pressure, the composition of matter is changing in a
sequence of nuclear reactions: electron captures, neutron emission and absorption, and
finally, at densities exceeding $10^{12}~{\rm g~cm^{-3}}$, also pycnonuclear fusion.
As the nuclear processes are  proceeding off-equilibrium, they are accompanied by the
heat deposition in the crustal matter. For the HZ model, the total crustal heating
amounts to $1.4~$MeV per one accreted nucleon. This heating is mostly supplied by the
pycnonuclear fusion processes in the inner crust at $\rho=10^{12}-10^{13}~{\rm
g~cm^{-3}}$.

Very recently, new X-ray burst simulations, with a network of nuclei much larger than
the previously used networks which ended with the iron-peak nuclei,  were carried out
by Schatz et al. (2001). Extending the network of possible nuclear reactions resulted
in nuclear ashes  composed predominantly of nuclei much heavier than iron, with $A$ up
to $\sim 112$. The question arises how changing the initial composition from
$^{56}{\rm Fe}$ to, e.g., $^{104}{\rm Te}$ will change the crustal heating  and the
composition of the deeper layers of the inner crust. The present Letter is devoted to
answering this question.

Brief description of the dense-matter model and of the processes taking place in an
accreting neutron-star crust is given in Sect.\ 2. Results for the composition and
heating, obtained for various initial compositions and dense-matter models, are
presented in Sect.\ 3. Final Sect.\ 4 is devoted to a discussion of our results and
conclusion.
\section{Non-equilibrium nuclear processes}
\label{sect:nuc.acc}
In what follows we briefly describe the nuclear evolution scenario
of HZ as applied in the present study (earlier studies on various
aspects of the non-equilibrium neutron-star crusts were done by
Vartanyan \& Ovakimova 1976, Bisnovatyi-Kogan \& Chechetkin 1978,
Sato 1979). Under conditions prevailing in accreting neutron-star
crust at $\rho>10^8~{\rm g~cm^{-3}}$ matter is strongly
degenerate, and is relatively cold ($T < 10^8~{\rm K}$), so that
thermonuclear processes involving charged particles are blocked.
At the densities lower than the threshold for the pycnonuclear
fusion $\rho_{\rm pyc}\sim 10^{12}~{\rm g~cm^{-3}}$, the number of
nuclei in an element of matter does not change during the
compression resulting from the increasing weight of accreted
matter. For simplicity, we assume that only one nuclear species
$(A,Z)$ is present at each pressure. Due to the nucleon pairing,
stable nuclei in dense matter  have even $N=A-Z$ and $Z$
(even-even nuclides). In the outer crust, in which free neutrons are
absent, the electron captures which proceed in two steps,
\begin{eqnarray}
(A,Z)+e^-&\longrightarrow & (A,Z-1)+\nu_e~, \cr\cr (A,Z-1)+e^-&\longrightarrow &
(A,Z-2)+\nu_e + Q_{\rm c}~. \label{eq:e.cap}
\end{eqnarray}
lead to a systematic decrease of $Z$ (and increase of $N=A-Z$)
with increasing density.  The first capture in Eq.\
(\ref{eq:e.cap}) proceeds in a quasi-equilibrium manner, with a
negligible energy release. It produces an odd-odd nucleus, which
is strongly unstable in dense medium, and captures a second
electron in an non-equilibrium manner, with energy release $Q_{\rm
c}$. A fraction of  this energy, $Q_{\rm d}$, is deposited in the
matter and heats it, but most of it, $Q_\nu$, is carried away by
the neutrino,  so that $Q_{\rm c}=Q_{\rm d}+Q_\nu$. The effective
heat deposited in matter is
\begin{equation}
Q_{\rm d}=\eta (\mu_e -\Delta) + Q_{\rm exc}~,
\label{eq:Q.dep.cap}
\end{equation}
where $\mu_e$ is the electron Fermi energy (including the rest energy)
 and $\Delta$ is the energy threshold for the first quasi-equilibrium electron capture.
 $Q_{\rm exc}$ is the excitation energy of the final nucleus , and  $\eta$ ranging from $1/6$
 to $1/4$
 accounts for the neutrino energy losses.
 We put $Q_{\rm exc}=0$ in the actual simulations, except for
one case with initial $^{56}{\rm Fe}$ composition when $Q_{\rm exc}$
is experimentally known (see HZ).
 We have $\eta=1/6$ for $\mu_e\gg \Delta$ and
 $\eta=1/4$ for $(\mu_e-\Delta)/\mu_e\ll 1$.
 In the crust evolution simulation the latter condition is
fulfilled and therefore we put $\eta=1/4$.
  Above the neutron-drip point
 ($\rho>\rho_{\rm ND}$), electron captures trigger neutron emissions,
\begin{eqnarray}
(A,Z)+e^-&\longrightarrow & (A,Z-1)+\nu_e~, \cr\cr (A,Z-1)+e^-&\longrightarrow &
(A-{\rm k},Z-2)+{\rm k}\; n + \nu_e + Q_{\rm c}~, \label{eq:e.cap.n}
\end{eqnarray}
where the number of emitted neutrons ``k'' is even. Due to the electron captures, the
value of $Z$ decreases with increasing density. In consequence, the Coulomb barrier
prohibiting the nucleus-nucleus reaction lowers. This effect, combined with the
decrease of the separation between the neighboring nuclei, and a simultaneous increase
of energy of the quantum zero-point vibrations around the nuclear lattice sites, opens
a possibility of the pycnonuclear  reactions. The pycnonuclear fusion timescale
$\tau_{\rm pyc}$ is a very sensitive function of $Z$. The chain of the reactions
(\ref{eq:e.cap.n}) leads to an abrupt decrease of
$\tau_{\rm pyc}$ typically by 7 to 10 orders of magnitude.
Pycnonuclear fusion switches-on as
soon as $\tau_{\rm pyc}$ is smaller than the time of the
 travel  of a piece of matter (due to the accretion) through the considered
shell of mass $M_{\rm shell}(N,Z)$,
$\tau_{\rm acc}\equiv M_{\rm shell}/\dot{M}$. The masses of the shells are
of the order of $10^{-5}~ {\rm
M}_\odot$. As a result the point where the pycnonuclear
reaction takes place is very well defined and  the
chain of reactions (\ref{eq:e.cap.n}) in several cases is
followed by the pycnonuclear reaction on a timescale
much shorter than $\tau_{\rm acc}$.
Denoting $Z'=Z-2$, we have then
\begin{eqnarray}
&~&~~~~~~~~(A,Z')+(A,Z')\longrightarrow  (2A,2Z')+Q_1~,\cr\cr &~&(2A,2Z')\longrightarrow
(2A-{\rm k'}, 2Z')+{\rm k'}\;n + Q_2~,\cr\cr &~&\ldots~~~~~~\ldots~~~~~~\ldots~~~~~~\ldots +Q_3~,
\label{eq:pyc.scheme}
\end{eqnarray}
where ``$\dots$'' correspond to some not specified chain of the
electron captures accompanied by neutron emissions. The total heat
deposition in matter, resulting from a chain of reactions
involving a pycnonuclear fusion, is $Q_{\rm pyc}=Q_1+Q_2+Q_{{\rm
3,d}}$ where label "d" indicates that only the fraction
deposited in the matter is included.

\begin{figure}
\resizebox{\hsize}{!}{\includegraphics[angle=0]{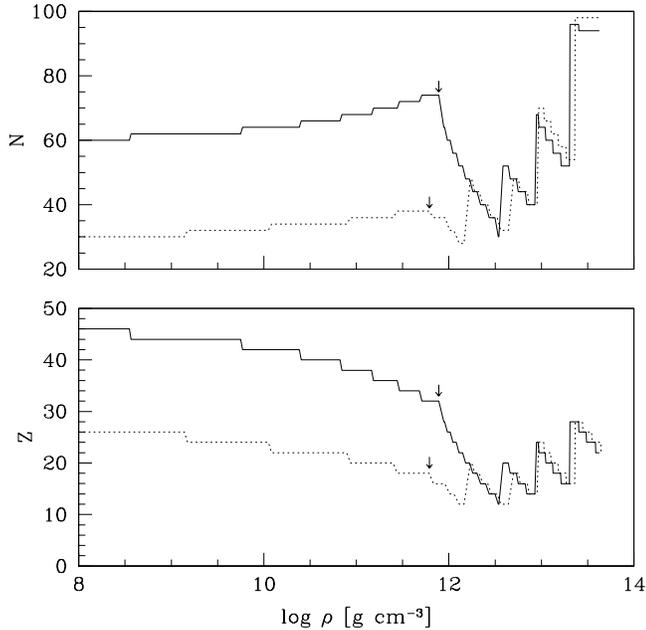}}
\caption{$Z$ and $N$ of nuclei versus matter density in an
accreting neutron-star crust. Solid line: $A_{\rm i}=106$; dotted
line: $A_{\rm i}=56$.  Each change of $N$ and $Z$, which takes
place at a constant pressure, is accompanied by a jump in density
(see HZ for detailed discussion of this point). Small steep
segments connect the top and the bottom density of thin reaction
shell. Arrows indicate positions of the neutron drip point. }
 \label{fig:ZNrho}
\end{figure}

Our model of atomic nuclei is described in HZ. Using our nuclear-evolution code, we
evolved an element of matter  consisting initially of nuclei $(A_{\rm i},Z_{\rm i})$
immersed in an electron gas, from $\rho_{\rm i}= 10^8~{\rm g~cm^{-3}}$ to
$\rho=\rho_{\rm f}>5\times 10^{13}~{\rm g~cm^{-3}}$. Our results for the composition and
crustal heating are presented in the next section.
\section{Composition and heating}
\label{sect:compos.heat}
The compositions of accreted neutron star crusts are shown in Fig. \ref{fig:ZNrho} and
in Table \ref{tab:crust.inner}.
These results describe  crusts built of accreted and processed matter up to the
density $5\times 10^{13}~{\rm g~cm^{-3}}$.
  At a constant accretion rate $\dot{M}=\dot{M}_{-9}\times 10^{-9}~{\rm
M}_\odot/{\rm yr}$ this will take $\sim 10^6~{\rm yr}/\dot{M}_{-9}$. During this
time, a shell of X-ray burst ashes will be compressed from $\sim 10^8~{\rm g~cm^{-3}}$
to $\sim 10^{13}~{\rm g~cm^{-3}}$.  Two different compositions of X-ray bursts ashes
at $\la 10^8~ {\rm g~cm^{-3}}$ were assumed. In the first case, $A_{\rm i}=56$, and
the HZ scenario is reproduced. In the second case, we consider an example of nuclear
ashes obtained by Schatz et al. (2001). To be specific, we assume $A_{\rm i}=106$. The
value of $Z_{\rm i}=46$ stems then from the condition of beta equilibrium at
$\rho=10^{8}~{\rm g~cm^{-3}}$. The compositions in the outer crust, where the only
processes are the electron captures, are strongly influenced by the initial
conditions. Up to the neutron-drip point, the difference by a factor of about two
between the values of $Z$ and $N$ for $A_{\rm i}=106$ and $A_{\rm i}=56$ is conserved.
It should be noted that in the case $A_{\rm i}=106$ the number of
beta captures in the outer crust is about 2 times larger, but each
reaction is accompanied by the density jump $\sim 5\%$, about half
of these in the case $A_{\rm i}=56$ (we have similar situation for the
energy release in a single shell).
One notices a dramatic effect of the neutron drip triggered by an electron capture at
$\rho=\rho_{\rm ND}$. We get $\rho_{\rm ND}=6\times 10^{11}~{\rm g~cm^{-3}}$ for
$A_{\rm i}=56$ and $\rho_{\rm ND}=8\times 10^{11}~{\rm g~cm^{-3}}$ for $A_{\rm
i}=106$.  Just after $\rho_{\rm ND}$ both $N$ and $Z$ of nuclei decrease in a long chain
of the neutron emissions followed by the electron captures.
 After the pycnonuclear fusion is switched-on at $\rho_{\rm pyc}\simeq
10^{12}~{\rm g~cm^{-3}}$, the two compositions converge, and stay very close,  up to
the largest densities beyond which the validity of the HZ model becomes questionable.
We checked that this is  a generic property of the $Z,N$ evolution, which does not
depend on specific values of $A_{\rm i},Z_{\rm i}$, or on the details of the nuclear
model used (see Sect.\ \ref{sect:conclusion}).


\newcommand {\st}{\rightarrow}
\begin{table*}[t]
\caption{
Non-equilibrium processes in the crust of an accreting neutron stars assuming
that the X-ray ashes consist of $^{106}{\rm Pd}$.
$P$ and $\rho$ are pressure and density at which the reaction
takes place.  $\Delta \rho/\rho$ is relative density jump
connected with reaction, $q$ is the heat deposited in matter,
$X_n$ is the fraction of free neutrons
among  nucleons, in the layer just above the reaction surface.
}
\label{tab:crust.inner}
\begin{center}
\begin{tabular}{llllrr}
\hline\hline
$P$ & $\rho$ & reactions & $X_n$  & $\Delta \rho/\rho$
& $q$\\
  (dyn~cm$^{-2}$)  &   (g~cm$^{-3}$)& & &\% & (keV)\\
\hline
 $  9.235\times 10^{25}$ &  $ 3.517\times 10^{08}$ &   $^{106}{\rm Pd}\st ^{106}{\rm Ru}-2e^-+2\nu_e $ & $     0      $  &$  4.4  $&$    5.7$ \\
 $  3.603\times 10^{27}$ &  $ 5.621\times 10^{09}$ &   $^{106}{\rm Ru}\st ^{106}{\rm Mo}-2e^-+2\nu_e $ & $     0      $  &$  4.6  $&$    5.7$  \\
 $  2.372\times 10^{28}$ &  $ 2.413\times 10^{10}$ &   $^{106}{\rm Mo}\st ^{106}{\rm Zr}-2e^-+2\nu_e $ & $     0      $  &$  4.9  $&$    5.6$  \\
 $  8.581\times 10^{28}$ &  $ 6.639\times 10^{10}$ &   $^{106}{\rm Zr}\st ^{106}{\rm Sr}-2e^-+2\nu_e $ & $     0      $  &$  5.1  $&$    5.6$  \\
 $  2.283\times 10^{29}$ &  $ 1.455\times 10^{11}$ &   $^{106}{\rm Sr}\st ^{106}{\rm Kr}-2e^-+2\nu_e $ & $     0      $  &$  5.4  $&$    5.5$  \\
 $  5.025\times 10^{29}$ &  $ 2.774\times 10^{11}$ &   $^{106}{\rm Kr}\st ^{106}{\rm Se}-2e^-+2\nu_e $ & $     0      $  &$  5.7  $&$    5.5$  \\
 $  9.713\times 10^{29}$ &  $ 4.811\times 10^{11}$ &   $^{106}{\rm Se}\st ^{106}{\rm Ge}-2e^-+2\nu_e $ & $     0      $  &$  6.1  $&$    5.5$  \\
 $  1.703\times 10^{30}$ &  $ 7.785\times 10^{11}$ &   $^{106}{\rm Ge}\st ^{92}{\rm Ni}+14n-4e^-+4\nu_e $ & $  0.13 $  &$ 13.2  $&$   77.6$  \\
 $  1.748\times 10^{30}$ &  $ 8.989\times 10^{11}$ &   $ ^{92}{\rm Ni}\st ^{86}{\rm Fe}+ 6n-2e^-+2\nu_e $ & $  0.19 $  &$  6.9  $&$   39.2$  \\
 $  1.924\times 10^{30}$ &  $ 1.032\times 10^{12}$ &   $ ^{86}{\rm Fe}\st ^{80}{\rm Cr}+ 6n-2e^-+2\nu_e $ & $  0.25 $  &$  7.3  $&$   43.1$  \\
 $  2.135\times 10^{30}$ &  $ 1.197\times 10^{12}$ &   $ ^{80}{\rm Cr}\st ^{74}{\rm Ti}+ 6n-2e^-+2\nu_e $ & $  0.30 $  &$  7.7  $&$   47.4$  \\
 $  2.394\times 10^{30}$ &  $ 1.403\times 10^{12}$ &   $ ^{74}{\rm Ti}\st ^{68}{\rm Ca}+ 6n-2e^-+2\nu_e $ & $  0.36 $  &$  8.1  $&$   52.3$  \\
 $  2.720\times 10^{30}$ &  $ 1.668\times 10^{12}$ &   $ ^{68}{\rm Ca}\st ^{62}{\rm Ar}+ 6n-2e^-+2\nu_e $ & $  0.42 $  &$  8.5  $&$   57.7$  \\
 $  3.145\times 10^{30}$ &  $ 2.016\times 10^{12}$ &   $ ^{62}{\rm Ar}\st ^{56}{\rm  S}+ 6n-2e^-+2\nu_e $ & $  0.47 $  &$  9.0  $&$   63.7$  \\
 $  3.723\times 10^{30}$ &  $ 2.488\times 10^{12}$ &   $ ^{56}{\rm  S}\st ^{50}{\rm Si}+ 6n-2e^-+2\nu_e $ & $  0.53 $  &$  9.4  $&$   70.5$  \\
 $  4.549\times 10^{30}$ &  $ 3.153\times 10^{12}$ &   $ ^{50}{\rm Si}\st ^{42}{\rm Mg}+ 8n-2e^-+2\nu_e $ & $  0.61 $  &$  8.8  $&$   79.0$  \\
\hline
 $  4.624\times 10^{30}$ &  $ 3.472\times 10^{12}$ &   $ ^{42}{\rm Mg}\st ^{36}{\rm Ne}+ 6n-2e^-+2\nu_e $  &&& \\
&&$^{36}{\rm Ne}+^{36}{\rm Ne}\st ^{72}{\rm Ca}$& $  0.66 $  &$   10.6  $&$    251.8$\\
\hline
 $  5.584\times 10^{30}$ &  $ 4.399\times 10^{12}$ &   $ ^{72}{\rm Ca}\st ^{66}{\rm Ar}+ 6n-2e^-+2\nu_e $ & $  0.69 $  &$  4.8  $&$   25.3$  \\
 $  6.883\times 10^{30}$ &  $ 5.355\times 10^{12}$ &   $ ^{66}{\rm Ar}\st ^{60}{\rm  S}+ 6n-2e^-+2\nu_e $ & $  0.72 $  &$  4.7  $&$   27.3$  \\
 $  8.749\times 10^{30}$ &  $ 6.655\times 10^{12}$ &   $ ^{60}{\rm  S}\st ^{54}{\rm Si}+ 6n-2e^-+2\nu_e $ & $  0.75 $  &$  4.6  $&$   29.2$  \\
\hline
 $  1.157\times 10^{31}$ &  $ 8.487\times 10^{12}$ &   $ ^{54}{\rm Si}\st ^{46}{\rm Mg}+ 8n-2e^-+2\nu_e $ &&&  \\
&&$^{46}{\rm Mg}+^{46}{\rm Mg}\st ^{92}{\rm Cr}$& $   0.79 $  &$  4.0  $&$    139.6$\\
\hline
 $  1.234\times 10^{31}$ &  $ 9.242\times 10^{12}$ &   $ ^{92}{\rm Cr}\st ^{86}{\rm Ti}+ 6n-2e^-+2\nu_e $ & $  0.80 $  &$  2.0  $&$  8.9$  \\
 $  1.528\times 10^{31}$ &  $ 1.096\times 10^{13}$ &   $ ^{86}{\rm Ti}\st ^{80}{\rm Ca}+ 6n-2e^-+2\nu_e $ & $  0.82 $  &$  1.9  $&$  9.0$  \\
 $  1.933\times 10^{31}$ &  $ 1.317\times 10^{13}$ &   $ ^{80}{\rm Ca}\st ^{74}{\rm Ar}+ 6n-2e^-+2\nu_e $ & $  0.83 $  &$  1.8  $&$  8.8$  \\
 $  2.510\times 10^{31}$ &  $ 1.609\times 10^{13}$ &   $ ^{74}{\rm Ar}\st ^{68}{\rm  S}+ 6n-2e^-+2\nu_e $ & $  0.85 $  &$  1.7  $&$ 10.2$  \\
\hline
 $  3.363\times 10^{31}$ &  $ 2.003\times 10^{13}$ &   $ ^{68}{\rm  S}\st ^{62}{\rm Si}+ 6n-2e^-+2\nu_e $  &&&\\
&&$^{62}{\rm Si}+^{62}{\rm Si}\st ^{124}{\rm Ni}$& $  0.86 $  &$  1.7  $&$   70.3$ \\
\hline
 $  4.588\times 10^{31}$ &  $ 2.520\times 10^{13}$ &   $ ^{124}{\rm Ni}\st ^{120}{\rm Fe}+ 4n-2e^-+2\nu_e $ & $  0.87 $  &$ 0.8  $&$  2.6$  \\
 $  5.994\times 10^{31}$ &  $ 3.044\times 10^{13}$ &   $ ^{120}{\rm Fe}\st ^{118}{\rm Cr}+ 2n-2e^-+2\nu_e $ & $  0.88 $  &$ 0.9  $&$  2.4$  \\
 $  8.408\times 10^{31}$ &  $ 3.844\times 10^{13}$ &   $ ^{118}{\rm Cr}\st ^{116}{\rm Ti}+ 2n-2e^-+2\nu_e $ & $  0.88 $  &$ 0.8  $&$  2.2$  \\
\hline
\hline
\end{tabular}
\end{center}
\end{table*}

\begin{figure}
\resizebox{\hsize}{!}{\includegraphics[angle=0]{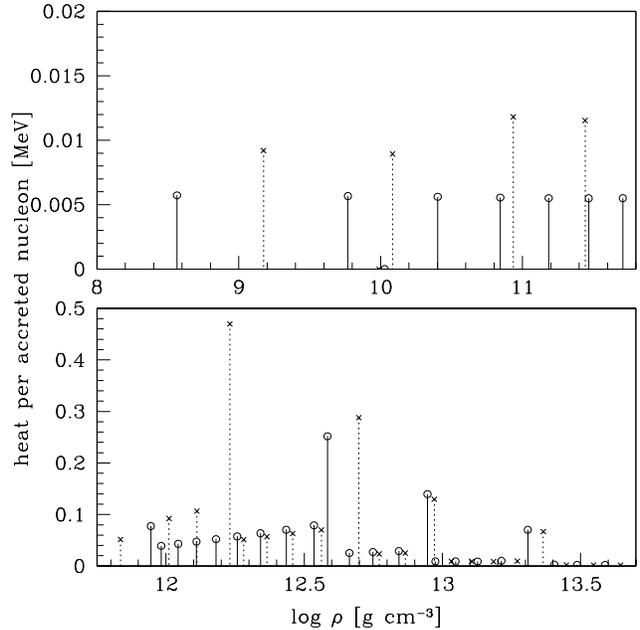}}
 \caption{Heat per one accreted nucleon, deposited in
the crust,  for two models with different initial $A$.
Solid vertical lines (ended with circles): $A_{\rm i}=106$;
dotted lines (ended with crosses): $A_{\rm i}=56$. Vertical lines are positioned
at the densities at
the bottom of the reaction shell.}
 \label{fig:Qrho}
\end{figure}

In Fig.\ \ref{fig:Qrho} we show the heat deposited in the matter, per one accreted
nucleon, in the thin shells in which non-equilibrium nuclear processes are taking
place. Actually, reactions proceed at a constant pressure, and there is a density jump
within a thin  ``reaction shell''. The vertical lines whose height gives  the heat
deposited in matter are drawn at the density at the  bottom  of the reaction shell.

%
One notices a specific dependence of the number of heat sources and the heating
power of a single source on assumed $A_{\rm i}$.
Let us start with the outer crust (upper panel of
 Fig.\ \ref{fig:Qrho}). In the case of $A_{\rm i}=56$ the number of sources is
 smaller, and their heat-per-nucleon values $q$ are larger, than for $A_{\rm i}=106$.
 Hovever, the total deposited heat-per-nucleon is quite similar,
 0.041 and 0.039 MeV/nucleon  for $A_{\rm i}=56$  and $A_{\rm i}=106$, respectively.
 Similar features are seen in the inner crust (lower panel of  Fig.\ \ref{fig:Qrho}).
 The total crustal heating is  1.54 MeV/nucleon  and 1.12 MeV/nucleon for
 $A_{\rm i}=56$ and $A_{\rm i}=106$, respectively. The difference
 $\simeq 0.4$~MeV/nucleon
 between these two cases is mainly due to an additional
 pycnonuclear reaction (the first one) in the case $A_{\rm i}=56$,
 which results in the convergence of two evolutionary scenarios at
 $\rho\simeq 2 \times 10^{12}~{\rm g~cm^{-3}}$. This pycnonuclear fusion
 is accompanied by the larger energy release than the subsequent beta captures
 and neutron emissions in the case $A_{\rm i}=106$.
 The nearly exact convergence of the cases $A_{\rm i}=56$ and $A_{\rm
 i}=106$ for $\rho > 10^{12}~{\rm g~cm^{-3}}$ is connected with
 the fact that heavier nucleus has $N_{\rm i}$ and $Z_{\rm i}$ which are nearly double of
 those of $^{56}{\rm Fe}$. In the case of the initial nuclei between $A_{\rm i}=56$
 and $A_{\rm i}=106$ the situation is similar with the nearly same
 nuclei after the first pycnonuclear reaction, with the slight  shift
 in the densities of the  boundaries between subsequent shells.
 The total energy per nucleon released then above the neutron
 drip point is between 1.1 and 1.5 MeV, with unchanged crustal heating below
 the neutron drip.
 The value obtained for $A_{\rm
 i}=56$ is slightly larger than that quoted in HZ, which results from  correcting
  too large neutrino losses in electron captures assumed in HZ.


\section{Discussion and conclusion}
\label{sect:conclusion}
We studied nuclear composition of accreted neutron-star crusts, assuming different
compositions of the X-ray bursts ashes. Obtained compositions ($A,Z$) of the outer
crust for ashes with $A_{\rm i}=56$ and $A_{\rm i}\simeq 100$ keep the difference by a
factor of about two up to the neutron-drip point. However, total crustal heating in the
outer crust is similar for both $A_{\rm i}$, and is negligible compared with the crustal
heating in the inner crust. A chain of processes occurring after the  neutron drip
leads to convergence of compositions to a common one, which at densities higher than
$10^{12}~{\rm g~cm^{-3}}$ does not depend on $A_{\rm i}$. While the number of the
heat-sources depends on $A_{\rm i}$, and is significantly larger for $A_{\rm i}\simeq
100$ than for $A_{\rm i}=56$, total crustal  heating, 1.1 MeV/nucleon and 1.5 MeV/nucleon, is
quite similar.  Simulations with other values of $A_{\rm i}$ confirm that the crustal
heating is only moderately dependent on the nuclear composition of the X-ray bursts
ashes.

Let us mention, that in a very recent paper Schatz et al.
(2003) suggest that rp-processes ashes can burn explosively at
density $\sim 10^9~{\rm g~cm^{-3}}$, giving rise to the X-ray
superburst. Such a burning will lead to photodisintegration of the
$A\sim 100$ nuclei and shift composition of the ashes
toward iron, so that in this case the original scenario of HZ
will hold.

In view of the importance of the crustal heating for the SXRTs
models, more detailed study of non-equilibrium processes in
accreting neutron-star crusts is desirable. An investigation along
these lines is now being carried out and its results will be
presented in a separate publication (Haensel and Zdunik, in
preparation).
\begin{acknowledgements}
 We are very grateful to the referee for detailed and helpful
comments.
We are also grateful to D.G. Yakovlev for useful correspondence.
This work was supported in part by the KBN grant 5 P03D 020 20.
\end{acknowledgements}

\end{document}